\documentclass{egpubl}
\usepackage{pgv2025}

\WsSubmission      %

\CGFccby

\usepackage{t1enc,dfadobe}

\BibtexOrBiblatex
\electronicVersion
\PrintedOrElectronic
\ifpdf \usepackage[pdftex]{graphicx} \pdfcompresslevel=9
\else \usepackage[dvips]{graphicx} \fi

\usepackage{egweblnk}

\usepackage{cite}  %
\usepackage[dvipsnames,svgnames]{xcolor}
\usepackage{soul}
\usepackage{cleveref}
\usepackage{booktabs} 
\usepackage{microtype} 

\usepackage[inkscapelatex=false]{svg}

\usepackage{hyperref}
\usepackage{algorithm}
\usepackage{algpseudocode}
\usepackage{listings}
\usepackage{tabu}
\usepackage{subcaption}

\usepackage{textcomp}                  %
\usepackage{mathptmx}                  %
\usepackage{amssymb}
\usepackage{times}                     %

\def\lstfontsize{\fontsize{7}{7}\selectfont}
\definecolor{codeGreen}{rgb}{0,0.6,0}
\definecolor{codeBlue}{rgb}{0,0,1}
\definecolor{codeRed}{rgb}{0.65,0.11,0.36}
\definecolor{codeGray}{rgb}{0.6,0.6,0.6}
\definecolor{codeMauve}{rgb}{0.58,0,0.82}
\definecolor{codeCyan}{rgb}{0,0.52,0.70}
\definecolor{codeOrange}{rgb}{0.9,0.65,0.0}

\newcommand{\rev}[1]{{\color{purple}#1}}
\newcommand{\revdel}[1]{{\color{purple}\st{#1}}}
\renewcommand{\rev}[1]{#1}
\renewcommand{\revdel}[1]{}

\title[A Cache-Accelerated INR framework for Interactive Visualization of Tera-Scale Data]%
      {From Cluster to Desktop: A Cache-Accelerated INR framework for Interactive Visualization of Tera-Scale Data}

\author[Zavorotny et. al.]
{\parbox{\textwidth}{\centering Daniel Zavorotny$^{1}$,
        Qi Wu$^{2}$,
        David Bauer$^1$,
        Kwan-Liu Ma$^{1}$
        }
        \\
{\parbox{\textwidth}{\centering $^1$University of California, Davis\\
         $^2$NVIDIA
       }
}
\vspace{-2.0em}
}

\begin{document}

\maketitle
\begin{abstract}
Machine learning has enabled the use of implicit neural representations (INRs) to efficiently compress and reconstruct massive scientific datasets. However, despite advances in fast INR rendering algorithms, INR-based rendering remains computationally expensive, as computing data values from an INR is significantly slower than reading them from GPU memory. This bottleneck currently restricts interactive INR visualization to professional workstations. To address this challenge, we introduce an INR rendering framework accelerated by a scalable, multi-resolution GPU cache capable of efficiently representing tera-scale datasets. By minimizing redundant data queries and prioritizing novel volume regions, our method reduces the number of INR computations per frame, achieving an average 5× speedup over the state-of-the-art INR rendering method while still maintaining high visualization quality. Coupled with existing hardware-accelerated INR compressors, our framework enables scientists to generate and compress massive datasets in situ on high-performance computing platforms and then interactively explore them on consumer-grade hardware post hoc.


\end{abstract}  

\section{Introduction}
The capabilities of modern scientific computation systems are increasing steadily with the rise of more powerful compute hardware and algorithm optimizations. In turn, the datasets that are produced by simulations and experiments also increase in size and complexity. In many cases, the sizes of these datasets now vastly out-scale the size of GPU memory or even system memory, rendering large data visualization impractical on consumer hardware. This high memory requirement has lead to increased research interest in machine learning methods for data compression. With these advancements in neural compression, scientists can train compact implicit neural representations (INRs) to replace datasets with explicit voxel values at render time. These INRs can be orders of magnitude smaller than their explicit counterparts, enabling easy transportation of these datasets as model parameters for downstream reconstruction and visualization. Specifically, the state-of-the-art INR method \cite{Wu:10175377} has already demonstrated the ability to achieve compression ratios exceeding 1000:1 on large datasets while often taking under a minute to train. While efficient in memory, interactive visualization using these INRs is similarly difficult to achieve on consumer grade workstations, because inferencing these models on tens of thousands of ray traced samples every frame is still very computationally expensive.

We make a key observation that current INR pipelines have to regenerate all sampled data each frame. This imposes a high computational load each frame as the same data is repeatedly generated, even at times where there is little change in scene parameters like the camera position or transfer function. We can address this issue by introducing a cache to store previously generated data. However, a simple LRU cache will quickly be out-scaled with increasing data size. Such a cache will only be able to represent a small portion of the visible data and may even degrade performance as the cache miss rate increases.

We develop a pipeline that enhances visualization interactivity in INR-based rendering by employing a scalable GPU cache. Our approach leverages the \textit{Multi-Level Multi-Resolution Page Table} (MRPD) caching method, originally proposed by Hadwiger et al.~\cite{Hadwiger} and later optimized by Sarton et al.~\cite{sarton:hal-01705431}. MRPD allows cache entries to cover larger regions of the volume by loading specific chunks or “bricks” of data at higher levels of detail (LoD). Combined with a cached page table hierarchy (discussed in detail in~\cref{sec:cache-arch}), this method enables efficient indexing and management of even peta-scale datasets. Our contribution first reconciles the benefits of INR compression with the performance and flexibility of a multi-resolution GPU cache, and then further improves MRPD by incorporating a saliency-based priority scheduling scheme.

To evaluate our approach, we conducted a series of experiments on various datasets to assess performance, quality, and memory efficiency. The test results indicate that our cache improves rendering performance by an average of 5$\times$ while maintaining quality comparable to the baseline. Furthermore, we present a case study where we compress a tera-scale dataset on a GPU cluster and demonstrate interactive rendering on a consumer grade system. To promote reproducibility and further research, our implementation will be publicly available at \url{https://github.com/VIDILabs/cINR}. 

The contributions of our work can be summarized as follows:
\begin{itemize}
    \item A full pipeline implementation and demonstration of tera-scale data compression utilizing an INR.  
    \item A rendering algorithm enabling interactive INR visualization of massive datasets on consumer grade hardware. 
    \item A cache scheduling method capable of prioritizing the most salient regions in the volume to maximize cache utilization.  
\end{itemize}

\begin{figure*}[tb]
    \centering
    \includegraphics[width=\linewidth]{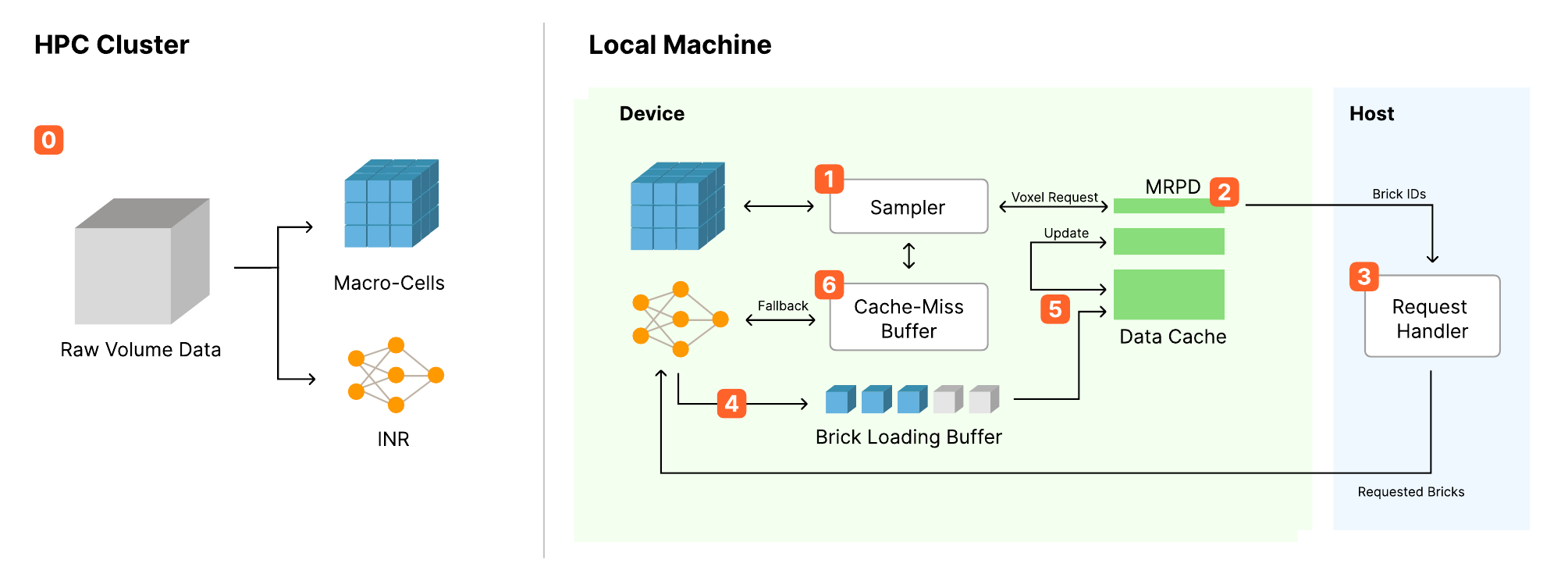}
    \vspace{-1em}
    \caption{After generating our macro-cell structure and compressed INR (0), our wavefront renderer utilizes the sampler interface (1) to request voxels from the MRPD cache manager (2). Missing brick ids are sent to the request handler in small batches (3) where they are scheduled for individual loading on the GPU (4) before being transferred into the data cache. The page table hierarchy and LRUs are then updated accordingly (5). Finally, missing voxels are inferred from the INR by the application (6) at the end of each sampling step.}
    \label{fig:overview}
    \vspace{-0.5em}
\end{figure*}

\section{Related Work}
In this section, we review related work in INR-based volume compression, inference optimization, and out-of-core streaming, highlighting key advancements and their relevance to our approach.

\textbf{INR Volume Compression.}
Traditional lossy compression algorithms such as \textit{TTHRESH} \cite{TTHRESH}, \textit{SZ}, \cite{sz} \textit{ZFP} \cite{zfp}, and \textit{SPERR} \cite{sperr} achieve impressive compression ratios for large floating-point datasets while maintaining high visual fidelity. However, these methods are not ideal for interactive visualization because individual samples or regions cannot be accessed without a broader decompression step. In contrast, implicit neural representations (INRs) eliminate the need for explicit decompression before sampling. Early work by Lu et al.~\cite{Lu} on \textit{neurcomp} combined SIREN-based~\cite{Sitzmann} INRs with ResNets~\cite{resnet} to achieve remarkable compression ratios and convincing reconstruction quality. Nevertheless, the high costs associated with model training and inference made this approach unsuitable for interactive visualization, even on high-end hardware. Later improvements by Weiss et al.~\cite{Weiss} and Wu et al.~\cite{Wu:10175377} reduced these costs by leveraging GPU tensor cores and employing trainable input encodings in the form of grids~\cite{Takikawa}  or hash tables~\cite{muller}. These advancements have reduced INR training times on most datasets from hours to seconds.

\textbf{INR Inference Optimizations.}
The greatest challenge for interactive INR-based rendering lies in the recurring cost of inferring many volume samples each frame. Calculating values through a fully connected multilayer perceptron (MLP) is substantially more resource-intensive than retrieving them directly from GPU memory. To bridge this gap, researchers have explored various strategies outside the realm of scientific visualization. For instance, Garbin et al.~\cite{Garbin} partitioned the Neural Radiance Field (NeRF)~\cite{Mildenhall} into separate components to independently learn coordinate and view direction representations, thereby reducing training complexity and overall model size. Similarly, DeRF~\cite{DeRF} and KiloNeRF~\cite{kilonerf} enhanced real-time NeRF rendering by assigning small, specialized networks to distinct volume segments. Further, MeRF~\cite{merf} improved memory efficiency with a sparse feature grid and 2D feature planes, enabling real-time rendering in web browsers, while SMERF~\cite{smerf} extended this approach with a hierarchical model partitioning scheme to support large-scale scenes. In our work, we tackle similar challenges within the context of scientific visualization.

\textbf{Out-of-Core Streaming.}
Our rendering method is inspired by techniques originally developed for out-of-core data streaming, which have become standard for real-time visualization of large-scale data. These techniques primarily address the challenge of managing I/O overhead when sampling from slower external memory. Early works by LaMar et al.~\cite{LaMar} and Zimmermann et al.~\cite{Zimmermann} used Level of Detail (LoD) sampling to minimize I/O operations. These methods were further refined on GPUs by Gobbetti et al.~\cite{Gobbetti} using octrees, with subsequent improvements by Crassin et al.~\cite{Crassin} through the use of GigaVoxels and view-dependent occlusion. Engel et al.~\cite{Engel} later adapted these techniques for tera-scale volume datasets~\rev{, while work by Lukas et al. \cite{residency} advanced techniques for multi-resolution rendering by using lower resolution bricks in place of missing cache entries}. Other spatial partitioning methods, such as KD-trees~\cite{Zellmann} and BSP trees~\cite{Xue}, have also been proposed to create tree-like file structures of multi-resolution volume bricks, often combined with an LRU cache on the GPU to reduce I/O during rendering. Hadwiger et al.~\cite{Hadwiger} advanced the field further by introducing a generalizable caching algorithm for tera-scale image stacks, utilizing a hierarchy of virtual memories---commonly known as the Multi-Resolution Page-Directory (MRPD) cache. This approach was refined by Sarton et al.~ \cite{sarton:hal-01705431} for rendering regular volumes and \rev{later adapted for} time-varying AMR grids~\cite{alexandre2023gpu}, and it forms the basis of our work.

\begin{figure}[h]
    \centering
    \includegraphics[width=0.9\linewidth]{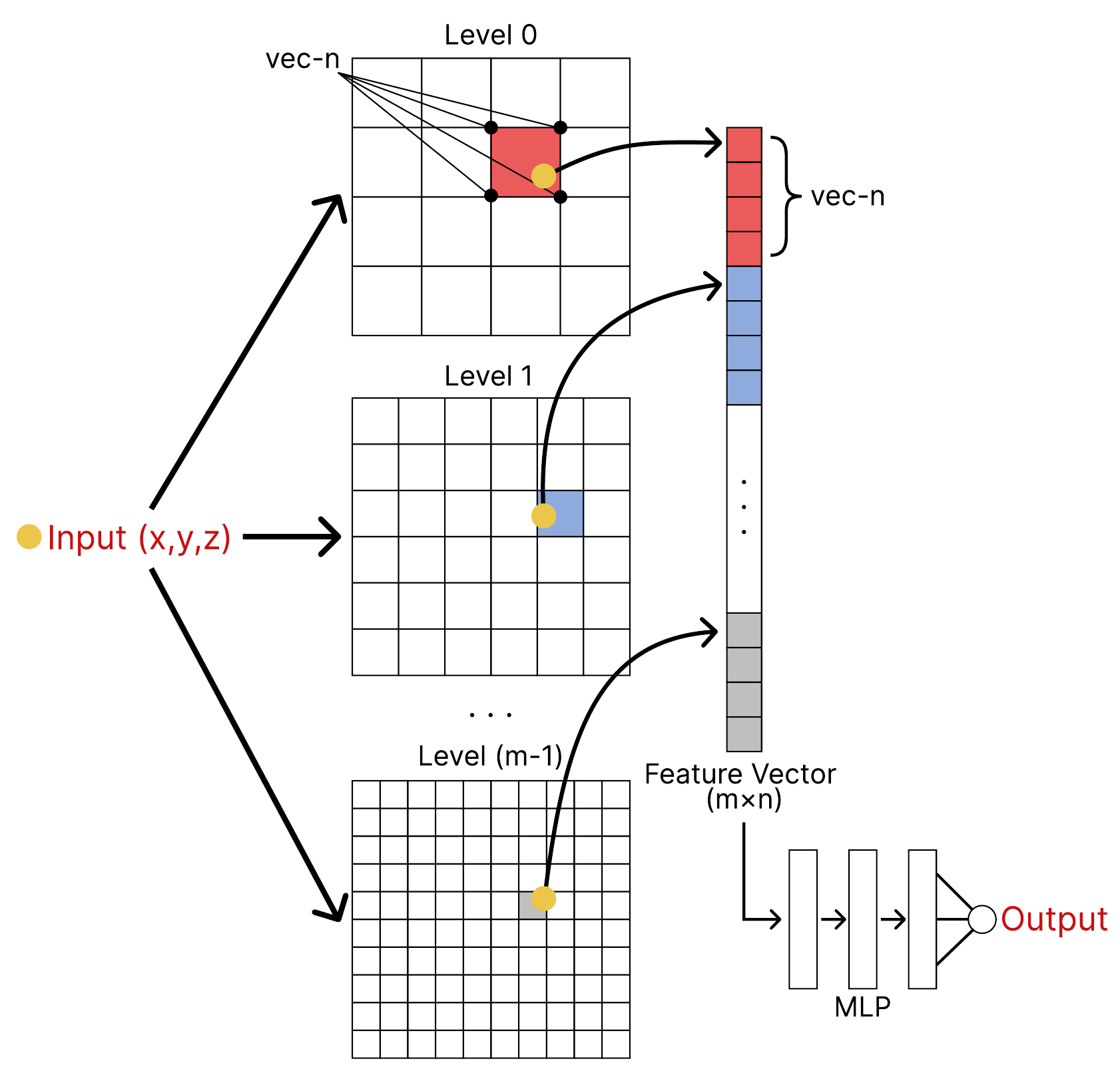}
    \vspace{-0.25em}
    \caption{An overview of the model architecture and hash-grid encoder. A vector of size $n$ is constructed from each resolution grid via interpolation of nearby grid points. The resulting feature vector thus encodes a multi-resolution representation of the input, allowing for a smaller MLP.}
    \label{fig:model-arch}
    \vspace{-0.5em}
\end{figure}

\section{Background}
Scientific volume data can be succinctly represented as a scalar field of the form $\Phi$: $\mathbb{R}^3 \to \mathbb{R}$, which maps a spatial coordinate $(x,y,z)\in \mathbb{R}^3$ to a discrete value $\textbf{v}\in \mathbb{R}$, representing the data entry at that spatial location. Overall, each voxel in the field can be sampled as $~(x,y,z) \mapsto \Phi(x,y,z) = \textbf{v}$. During visualization, each sampled voxel is passed through a transfer function which identifies its color and opacity. The static nature of this field allows for a single INR to optimize $\Phi$ per dataset, as opposed to NeRF~\cite{Mildenhall} or light field models which learn to directly reproduce lighting effects, requiring re-training for unique transfer functions. Additionally, since INRs are defined across the continuous domain $\mathbb{R}^3$, they can generate data samples at arbitrary resolutions without interpolation. 

In our work, the bulk of our model parameters is dedicated to a hash grid encoder \cite{muller}, adapted for scientific visualization by Wu et al. \cite{Wu:10175377}. This method, outlined in \cref{fig:model-arch}, defines $m$ 3D grids, each representing a different resolution level of the volume. Grid points of each resolution are stored using a fixed-sized hash table. Every table entry is associated with $n$ trainable parameters, representing local volume features at each respective level. For any given input coordinate, a feature vector is calculated for each grid resolution level by interpolating the nearby grid points. A final feature vector of size $n \times m$ is constructed via concatenation and then used as the input to the downstream fully-connected layers. This input encoder is able to capture high-frequency local features efficiently, making it possible to keep the MLP small and shallow. 

\section{INR framework for Interactive Visualization}
As scientific simulations and data gathering increase in scope, they demand high-end hardware not only to compute, but also for post hoc visualization. We identify memory, in particular, as the most significant limiting factor constraining these visualizations to high-end machines or GPU clusters. Out-of-core rendering algorithms require mass external storage to access the original data, and traditional compression techniques, while achieving great compression ratios and superior quality preservation, require an explicit decompression step to access the data. INRs, which may not reach the same qualities offered by more traditional compression algorithms, remain compact throughout rendering. INR models remain fully in-core while representing volumes that are orders of magnitude larger than GPU memory. In the following sections, we will give an overview of our entire data-to-visualization pipeline (see \Cref{fig:overview}) utilizing a current state-of-the-art INR model by Wu et al. \cite{Wu:10175377} which we enhance its interactivity with a scalable, multi-resolution GPU cache.

\subsection{Compression}
The compression stage of our pipeline is fairly straightforward as we just need to optimize on the function $\Phi$ for our chosen INR. We implement this stage following Wu et al.'s method \cite{Wu:10175377}. An important note is their observation that values with a high dynamic range over both coordinates and voxels can make training unstable and potentially impossible to optimize. Thus training is conducted over both the normalized domain $[0,1)^3$ and ground truth data $[0, 1]$. It is important to keep note of a dataset's exact value range during training for proper normalization at render time as well. 

During training, we decide between three separate pipelines depending on if the data can fit into total GPU memory, can fit into system memory, or requires mass storage. In the first two cases, we refine our parameters over a series of training steps in which we first generate a batch of random coordinates (batch size = 65536) with their respective ground truth values, and transfer the data onto the GPU, if not already loaded, for optimization. If the data is unable to fit even into system memory, we use Wu et al.'s out-of-core sampling method \cite{Wu:10175377} to maintain an asynchronously updated random data buffer in system memory, from which we sample instead. Each use case imposes its own cost which increases training time accordingly. However, with a sufficiently sized GPU cluster, we find that large volumes take only a few minutes minutes to complete training. Out-of-core sample streaming may take several hours depending on the speed of the external storage that houses the data.  

\subsection{Visualization}
The primary strength of INRs over traditional data compression methods is that they do not require an explicit decompression step. Traditional compression for post hoc scientific visualization results in a dynamic memory load as chunks of the data must be decompressed for each sampled region. Additionally, since INR features train on a continuous domain, the volume can be up-scaled or down-scaled at render time to arbitrary resolutions without interpolation artifacts affecting visual fidelity. On more computationally limited workstations, large volumes can be down-scaled to boost performance with minimal loss in quality. However, we use this flexibility in our results to up-scale existing volumes instead to make our smaller datasets more challenging to render.

\subsubsection{Rendering Challenges}
While memory and I/O are big advantages of INR rendering over out-of-core methods, INR models are not cheap to sample from. Each query of an INR has to be passed through the entire network. While inference is trivially parallelizable on GPUs, it is still easy to observe that a direct memory lookup would be faster. In addition, many existing INR rendering approaches discard queried values after a single use, leading to a very high inference overhead as many of the same coordinates will be inferred again in the following frames. A cache is a good way to reduce this overhead. A simple LRU cache can store previously sampled voxels and make them available in subsequent frames. However, a cache is also limited by the available GPU memory, and a na\"ive implementation may even deteriorate performance on large data due to the large amount of cache misses as the cache will only be able to represent a small portion of the volume.

\subsection{Accelerating Rendering with a GPU Cache}
In order for a cache to be useful on large data, it needs to minimize cache misses by representing a significant portion of data sampled each frame. A multi-resolution rendering approach gives the flexibility required to build such a cache, and Sarton et al.'s method \cite{sarton:hal-01705431} provides us with three core design decisions which we leverage to great effect in our work. First, data in the cache is stored in cubic chunks or ``bricks'', not as individual voxel-coordinate pairs. These bricks are all of the same size, but may represent much larger regions of the volume at a lower resolution, higher LoD. Next, whenever the sampler checks the cache for a given value, but the cache encounters a miss at the requested LoD, it will check if that value is represented by a brick at a higher LoD. Only if a given coordinate is not covered by any brick in that region will the cache signal a miss to the sampler. Internally however, a brick request will be raised for the original missing value. Finally, the cache schedules new bricks to be loaded asynchronously. In Sarton et al.'s \cite{sarton:hal-01705431} out-of-core implementation, this step was crucial for hiding I/O while loading from external memory, which is orders of magnitude slower than GPU memory. However, we find that even though our pipeline remains fully in-core, allowing this asynchronous scheduling removes jittering and smooths out interactive viewing. Aside from utilizing these features, we also enhance the way these requests are managed by internally ranking the requested bricks. We prioritize requesting the most impactful bricks first which gives us more stable performance across diverse data sets.

\subsubsection{Caching Architecture}
\label{sec:cache-arch}
One challenge of making a scalable cache architecture is managing entries. Each brick in the volume should have a unique identifier the cache can query against for its status. Storing these entries all on the GPU would present an issue as the number of bricks increases, especially considering that each LoD adds a set of unique bricks as well. We again look at Sarton et al.'s work \cite{sarton:hal-01705431}
for our design inspiration. Their work has the capability to fully address massive qualities of bricks by utilizing a separately cached, \textit{Multi-Level Multi-Resolution Page Table} (MRPD) hierarchy. This pyramidal structure contains an entry for each brick in the volume which flags to the cache if that brick is \textit{Mapped}, \textit{Unmapped}, or \textit{Empty}. \textit{Mapped} bricks are those currently in the cache and additionally record their location in the data cache. It follows then that \textit{Unmapped} bricks are not represented yet in the cache and sampled coordinates that land in them will trigger either a lookup in a higher LoD or a cache miss. Finally, \textit{Empty} entries are used to keep track of empty volume regions and immediately return a default value. We remove this flag from our design however, as we address empty space skipping differently in \cref{sec:adaptive-sampling}. For small volumes, the MRPD can directly store all entries on the GPU without any virtualized memory. On sufficiently large data, the cache dynamically adds levels of virtualized \textit{Page Tables} PT to the MRPD to address the increasing number of entries. These PT are independently cached by level on the GPU and allow the MRPD to remain compact while addressing all relevant bricks. In practice, only two such levels are needed to fully address petabyte scale volumes, so the overhead of traversing the MRPD is small. A more detailed overview of this structure along with a example of a voxel request being processed may be found in the original paper by Sarton et al. \cite{sarton:hal-01705431}.

Aside from the MRPD, which remains unchanged in our work, the caching pipeline can be broadly broken down into two main components; the \textit{Cache Manager} CM which manages the MRPD hierarchy and voxel requests, and the \textit{Request Handler} RH which runs asynchronously and manages brick loading from batches of requests sent to it from the CM. Our priority ranking implementation is housed in the CM (see \cref{sec:priority_ranking}), while we perform our INR integration in the RH. 
In Sarton et al.'s method \cite{sarton:hal-01705431}, a preprocessing step is conducted before rendering which generates a file structure of individual, compressed bricks from the volume sampled at each LoD. At the end of each frame, the CM generates a short list of requested bricks (50 by default) and sends them to the RH. Asynchronously, each request creates a file path to the requested brick, and each one is loaded, decompressed, and placed into system memory. Once all the requested bricks have been fetched, the CM will be notified and copy them into the GPU cache at the end of the next frame.  

\subsubsection{INR Integration}
\label{sec:INR-integration}

In our work, rather than generating a file path, loading, and decompressing, we use the information provided with each brick request to generate the appropriate coordinates and infer the brick from an INR. We treat our INR as a replacement to the preprocessed brick structure. Every request includes both the positional id of the brick, a 3D index, and the LoD that the brick represents ranging from 0 to the volume's $max\_lod$. To simplify our coordinate generation, we precompute all coordinates for brick $(0,0,0)$ beginning at the origin of the volume at LoD 0, the native resolution. This way, we just need to accurately offset the coordinates for each request and scale the distance between them according to the square of the LoD. It is helpful to point out that, due to the continuous domain of the INR, no interpolation is needed when we are, effectively, down-sampling the volume by scaling the distance between points this way. 

Consider an example of a request with $id: (0,1,2)$ at $LoD: 1$, with a brick side length of 40. The stride between coordinates in this brick will be $2^{LoD}=2^1=2$. There will be no offset along the $x$-axis. The $y$ values will be shifted by one LoD-scaled brick length $40 \times2^{LoD}-1=79$, we subtract 1 to account for a single voxel \textit{covering}, or overlap, between bricks on each dimension. Likewise, the $z$ coordinates will begin from an offset of $40 \times 2^{LoD}\times 2 -1 = 159$. The origin coordinate for this requested brick will then be $(0,79,159)$. From there, all the remaining brick coordinates will be offset in relation to each other and the LoD stride. All of these coordinates are computed and normalized in parallel before being passed into our INR to generate the brick data. This also demonstrates the scaling capabilities of a multi-resolution approach. Each brick stores exactly $40^3$ voxels, but a brick in the next higher level of the LoD hierarchy represents a region of $80^3$ voxels. Likewise, a brick in the next LoD will represent $160^3$ voxels. Additionally, since our pipeline is not attempting to load data externally, we can move the new brick buffer into the GPU. This way, we avoid the RAM-to-GPU copy operation once a set of requests is processed and keep our entire pipeline fully in-core on the GPU. 

\subsubsection{Priority Ranking}
\label{sec:priority_ranking}

\begin{figure}
    \centering
    \includegraphics[width=\linewidth]{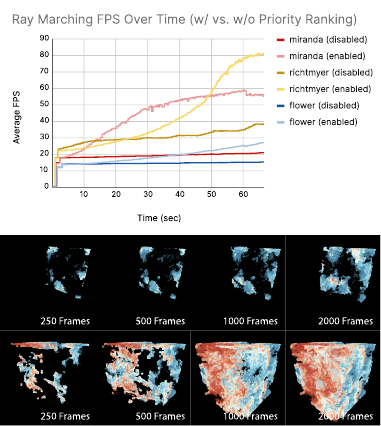}
    \caption{(Top) Performance comparison with our priority ranking enabled/disabled. (Bottom) Shows a timeline of the cache content after rendering without LoD pre-loading and fallback network calls on cache misses at $250$, $500$, $1000$, and $2000$ frames respectively. We see that ranking enables a more context aware representation of the data in the cache. }
    \label{fig:priority_ranking}
\end{figure}

By default, the cache design outlined in Sarton et al.'s work \cite{sarton:hal-01705431} treats all bricks as of equal value from the perspective of the CM. Upon being notified of a missing brick, the CM updates that brick's entry in an internal buffer with a timestamp of the current frame. Once the RH is available, the CM prepares a list of requests by gathering the first $n$ bricks matching the current frame's timestamp. In practice, this leads to bricks earlier in the buffer to be prioritized for loading, regardless of \rev{their location relative to the viewer or} how sparse they might be. \rev{Due to how the bricks are indexed, the cache will always load bricks closer to the volume origin first.} \rev{This effect has a disproportionate impact on performance in datasets where this region is more sparse or further away from the viewer. A good illustrative example can be seen with our ``Miranda'' dataset in \cref{fig:priority_ranking} where the furthest regions of the volume are loaded first due to the viewing angle. } \rev{We see that} the rendering performance benefits are highly data-dependent with this scheduling strategy.

\begin{figure*}[h]
    \centering
    \includegraphics[width=0.8\linewidth]{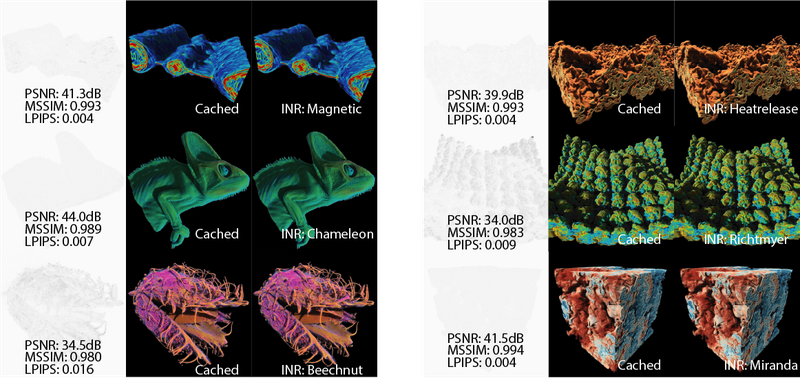}
    \caption{\rev{Image quality comparison of our multi-resolution pipeline, see ``Cached'' columns, with Wu et al.'s \cite{Wu:10175377} single resolution INR pipeline as our ground truth. The leftmost column of each comparison depicts pixel differences visually using FLIP \cite{flip}, lighter is better.} Additionally, we calculate PSNR, MSSIM, and LPIPs quality metrics based on both rendered images for each dataset. Results show that our cached pipeline maintains decent reconstruction quality while achieving the performance results shown in \cref{tab:1}.}
    \label{fig:quality}
    \vspace{-0.5em}
\end{figure*}

To address this issue, we \rev{develop} a technique which enhances the existing framework by ranking bricks by \rev{their} utilization. We keep the initial timestamp from when a missing brick was identified and increment its value by 1 with every subsequent request. Once the RH is prepared to accept a new request list, we gather the top $n$ bricks with the highest values. However, to also allow previously unseen bricks to be ranked fairly during interactive viewing, we clamp the max ranking at a value of 1000 over the timestamp and increment the timestamp accordingly each frame. We also find that updating the rank atomically is not needed since higher sampled regions will end up with an overall higher rank regardless of synchronicity. We find that fine ranking precision was not worth the trade-off.

Even a simple ranking operation like this greatly decreases the variance in cache performance per volume, and overall decreases the cache miss rate faster by prioritizing the most impactful bricks first. \rev{In \cref{fig:priority_ranking}, we showcase the improvements this brings to the performance of our cache. The figure also contains a visualization of how the cache fills up over time with and without ranking, showing the impact of this method on overall cache value distribution early on.}

\subsubsection{Handling Cache Misses}
The final component of our pipeline addresses true cache misses. That is, voxels that are not represented at any resolution in our cache. These are the most costly samples to deal with because we can only run a single instance of the INR on a given GPU. This slows down the rate at which our RH can processes brick requests as both resources utilize the same INR. 

We find that we can exploit the cache's own fallback procedure to minimize these true misses by pre-loading high LoD bricks. Our rendering application gives the user a variable to scale the distance at which samples are requested at higher LoDs. We begin rendering by forcing all samples to be at the highest LoD before smoothly transitioning to the user's chosen scale. This ensures that the cache contains its own, low resolution representation of nearly the entire volume to fall back on as the cache fills up. Pre-loading leaves much fewer true misses left for us to manage during the initial, and most costly, frames. Due to the misses being infrequent, we found it an acceptable approach to use an atomic function to record these coordinates directly as they come up (see \cref{fig:sampler-cache-code}). Once a sampling step is concluded, we infer all cache misses in a single pass through the INR.

\begin{figure}[h]
    \centering
    \includegraphics[width=\linewidth]{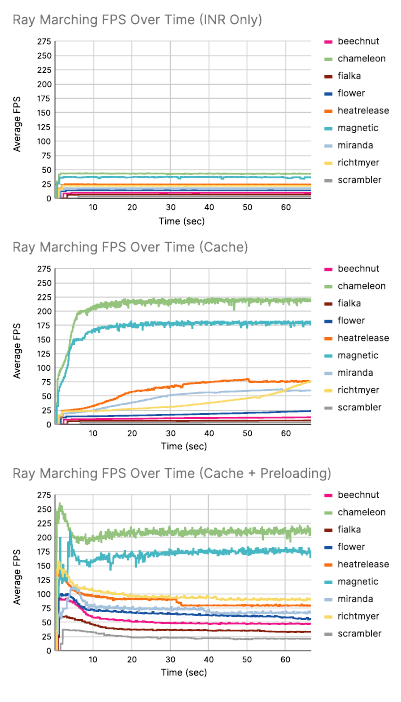}
    \vspace{-2.5em}
    \caption{FPS measured each frame over the course of our testing for \cref{tab:1}.
    (Top) Results with our cache disabled and sampling directly from the INR. (Middle) Results after enabling our cache without pre-loading higher LoDs. (Bottom) Results with pre-loading enabled. We see that pre-loading greatly improves performance during the initial frames and allows the FPS to stabilize quicker on the more challenging datasets. 
    }
    \label{fig:fps-over-time}
    \vspace{-0.5em}
\end{figure}

The effect that this method has on performance is shown in second and third plots in \cref{fig:fps-over-time}. Without pre-loading, our single INR is required to handle a massive amount of cache misses each frame, which slows down the rate the cache can be filled. This is especially pronounced in difficult datasets like \textit{flower} or \textit{beechnut}, which do not reach their maximum performance over the recorded time period. In contrast, performance with pre-loading is the highest initially before leveling off as more bricks of higher quality are loaded into the cache. 

\begin{algorithm}[tb]
\caption{\label{Algorithm:1}High-level sampling function which handles the pipeline's voxel requests, infers cache misses, and inserts missed values at the proper coordinates with minimal memory transfers.}%
\begin{algorithmic}
\Function{Sampler~}{\texttt{cache}, \texttt{positions}}
    \State \texttt{miss\_counter} $= 0$
    \State \texttt{missed\_coords$=$[\textit{empty list}]}
    \State \texttt{missed\_values$=$[\textit{empty list}]}
    \State 
    \State \texttt{SamplerKernel}(\texttt{miss\_counter}, \texttt{cache}, \texttt{positions})
    \If{\texttt{miss\_counter} $ > 0$}
        \State \texttt{InferINR}(\texttt{missed\_coords}, \texttt{missed\_values})
        \State \texttt{InsertMisses}(\texttt{missed\_coords}, \texttt{missed\_values})
    \EndIf{}
\EndFunction
\end{algorithmic}
\end{algorithm}

\begin{algorithm}[tb]
\caption{\label{fig:sampler-cache-code}Sampling kernel which queries the cache for requested voxels and atomically records cache misses in a global buffer to simplify handling and reinsertion.
}%

\begin{algorithmic}
\Function{SamplerKernel~}{miss\_counter, cache, position}
    \State \texttt{Dist$=$|position - camera\_pos|$\times$lod\_scale}
    \State 
    \State \texttt{LoD}$=$ \textit{integer part of} \texttt{Dist} \Comment{Stochastic LoD update}
    \If{\texttt{xorshift(Dist)$>$} \textit{decimal part of} \texttt{Dist}} 
        \State \texttt{LoD$=$LoD$+1$}
    \EndIf{}
    \State
    \State \texttt{Status$=$cache$\rightarrow$get\_sample(LoD, position)}
    \If{\texttt{Status$=$Hit}}
        \State \texttt{Output$=$} \textit{cached value}\Comment{Record the output}
    \EndIf{} \Comment{Record miss locations externally}
    \If{\texttt{Status$=$Miss}}
        \State \texttt{Output$=0$}
        \State \texttt{attomicAdd(miss\_counter$+1$)}
        \State \texttt{missed\_coords[miss\_counter]$=$position}
    \EndIf{}
\EndFunction
\end{algorithmic}
\end{algorithm}

\section{Rendering}
\rev{We base our approach on the state-of-the-art method for INR rendering~\cite{Wu:10175377}. This method is outlined in both \cref{sec:wavefront} and \cref{sec:adaptive-sampling}, which provide a brief overview of the rendering pipeline prior to the addition of our cache. We use this as the baseline for our INR results, with \cref{sec:stochastic-lod} covering our simple strategy to remove LoD boundary artifacts after enabling our cache.} A more detailed overview of the base rendering pipeline is presented in Wu et al's work \cite{Wu:10175377}. We extend their approach by introducing our multi-resolution cache and miss handling into their sampler in lieu of a direct INR inference, while keeping the rest of the pipeline unchanged.

\subsection{Wavefront Rendering}
\label{sec:wavefront}
In line with current state-of-art ray tracing practices, both our ray marching and path tracing pipelines follow a Wavefront \cite{wavefront} design architecture which decomposes a ray tracing renderer into distinct stages. For ray marching, we employ three separate kernels in a loop: a ray-generation kernel, coordinate-computation kernel, and the shading kernel. The ray-generation kernel initializes all primary rays, one per pixel, and calculates an intersection with the volume. Rays that do not intersect with the volume are pruned at this stage, and stream compaction is used to reduce the subsequent workload size. The coordinate-computation kernel then generates samples along each valid intersecting ray using our sampler kernel (\cref{fig:sampler-cache-code}). After the samples are generated, the identified cache misses are inferred and put back into place by our sampler (\cref{Algorithm:1}). Finally, the shading kernel reads the sampled values and calculates their impact on each ray's final color. This process continues in a loop until all rays are terminated and the final colors may be calculated. A similar strategy is used to decompose the path tracing pipeline utilizing Woodcock delta-tracking \cite{woodcock}. Both pipelines utilize the same sampling function, so no special considerations are needed for our cache.

\subsection{Adaptive Sampling}
\label{sec:adaptive-sampling}

Sarton et al.'s cache \cite{sarton:hal-01705431} relied on a pre-processing step to identify empty bricks in the volume. All entries in the MRPD corresponding to these bricks would then be marked as \textit{Empty}, and any samples mapping into these regions would immediately return some default value. Since our design removes this step, we need to turn to a different solution to address empty space skipping. We turn again to Wu et al.'s \cite{Wu:10175377} pipeline, which implements an adaptive sampling technique utilizing a light-weight, macro-cell acceleration structure generated during INR training. During this process, the volume is divided into 
${\lceil} \frac{V_x}{N} {\rceil}$
$\times$
${\lceil} \frac{V_y}{N} {\rceil}$
$\times$
${\lceil} \frac{V_z}{N} {\rceil}$
cubic chunks or "macro-cells" where $(V_x, V_y, V_z)$ represent the volume dimensions and $N$ each cell's side length. Each cell stores the minimum and maximum data value in the volume region it represents. Each time a transfer function is loaded by the application, these value ranges are used to determine the maximum opacity $\mu_{\max}$ for each cell, which then informs the sample frequency of each ray traveling through that cell. This structure allows the rays to entirely skip regions which are transparent in the current transfer function and adjust sampling rates in all other regions up or down, according to the $\mu_{\max}$. Additionally, unlike the MRPD's binary \textit{Empty} flag, this adaptive sampling approach gives us finer control over the sampling rate across the entire volume. 

\subsection{Stochastic LoD}
\label{sec:stochastic-lod}
We determine which LoD is most appropriate to sample from the cache based on both its distance from the camera, and a user-controlled scaling factor. However, both of these values are static within each frame and will cause all rays to transition between LoDs at the same distance from the camera, resulting in ring-like artifacts on the borders of these LoD boundaries. To improve visual quality, we implement stochastic LoD transitions for each ray to smooth out these boundaries. However, using a high quality random generator to vary these distances between rays will lead to a noticeable amount of overhead and affect performance. Therefore, we opted for a lighter, pseudo-random approach by employing a simple xorshift operation seeded with a higher-quality random value at the beginning of each frame. This operation is light enough to remain fully in situ in the sampler kernel (\cref{fig:sampler-cache-code}), and completely removes the visible artifacts between LoD boundaries.

\begin{table}[tb]
\centering%
\label{tab:1}
\resizebox{\linewidth}{!}{  
\begin{tabular}{rc|cc|cc}
    \toprule
    \multicolumn{2}{c|}{} & \multicolumn{2}{c}{Cached NGP} 
    & \multicolumn{2}{c}{Instant NGP} \\
    Dataset  & Dimensions & RM FPS & PT FPS & RM FPS & PT FPS \\
    \midrule
    Magnetic    & (2048, 2048, 2048) & 174.8 & 9.4  & 36.3 & 5.5  \\
    Chameleon   & (2048, 2048, 2160) & 212.3 & 20.6 & 43.0 & 11.0 \\
    Beechnut    & (2048, 2048, 3092) & 47.3  & 11.7 & 9.5  & 5.3  \\
    Fialka      & (3272, 3786, 1986) & 33.2  & 5.3  & 6.6  & 2.5  \\
    Flower      & (3652, 3234, 3828) & 56.4  & 1.7  & 14.0 & 0.8  \\
    Heatrelease & (4608, 1280, 3412) & 79.9  & 4.6  & 24.3 & 3.7  \\
    Scrambler   & (4354, 3870, 2612) & 21.1  & 2.9  & 3.3  & 1.2  \\
    Richmyer    & (4096, 4096, 3840) & 90.3  & 6.4  & 21.9 & 2.1  \\
    Miranda     & (4096, 4096, 4096) & 67.4  & 13.1 & 17.9 & 8.9  \\
    \textbf{DNS}& \textbf{(10240, 7680, 1536)} & 
    \textbf{91.8}  & \textbf{20.5} & \textbf{14.3} & \textbf{10.6} \\
    \bottomrule
\end{tabular}
}
\caption{Comparison of our cached pipeline's rendering performance to an un-cached method utilizing the same ray marching (RM) and path tracing (PT) algorithms. Our INR model architecture is fixed at 40.6MB for all datasets apart from DNS in which we use a larger network to represent the data more accurately at 150MB. For all experiments, we use the same cubic brick size of 40 voxels and a cache size of $30\times30\times30$ bricks. All experiments result in a similar VRAM utilization with a maximum of 11.3GB for DNS.}
\end{table}

\section{Results}
\label{sec:results}

We evaluated our implementation and collected all finalized results on a Linux machine with an NVIDIA RTX 3080Ti with 12 GB VRAM, an Intel 11700K, and 32 GB of DDR4 RAM. We used several up-scaled INRs representing datasets with different types of textures and challenges for rendering, a factor of 4$\times$ on each dimension for both \textit{Magnetic} and \textit{Miranda} and 2$\times$ for the rest. The rendered volume sizes of these INRs range from 34GB to 275GB for our up-scaled data, while our demonstration on the tera-scale DNS \cite{dns} dataset remains at native resolution for both compression and rendering. For consistency, we use the same brick size and cache dimensions for each test to maximize the utilization of our single GPU system. We up-scaled the smaller datasets to ensure that our cache would not be able to represent the scene entirely and our pipeline will always have cache misses to handle. \rev{We compare our results to a baseline method. To this end, we implement Wu et al.'s~\cite{Wu:10175377} INR renderer which is the state-of-the-art in INR rendering.}

\subsection{Rendering Performance}
To standardize measurements, all INRs are rendered for exactly 4000 frames. Our reported FPS metrics in \hyperref[tab:1]{Table 1} are an average FPS measurement over the last 200 frames generated for each volume in both Ray Marching (RM) and Path Tracing (PT) modes. We measure only on these last frames to insure the FPS is representative of the volume after the effects of pre-loading have disappeared. We observe an average 5x performance gain over pure INR rendering in RM mode, and a roughly 2x gain for PT on our workstation. The pipeline for PT itself is much more computationally expensive than RM, with the overhead of sample generation representing a proportionally smaller cost than RM. Additionally, PT samples areas of the volume out of view to calculate accurate shadows, resulting in more cache misses. Overall however, our results demonstrate smoother and more interactive performance over the standalone INR renderer. The two most challenging datasets we look at, \textit{Fialka} and \textit{Scrambler} render at less than 10 fps off the network alone, leaving them borderline non-interactive without our cache.

\begin{figure}[h]
    \centering
    \begin{subfigure}[t]{1.0\linewidth}
        \includegraphics[width=\linewidth]{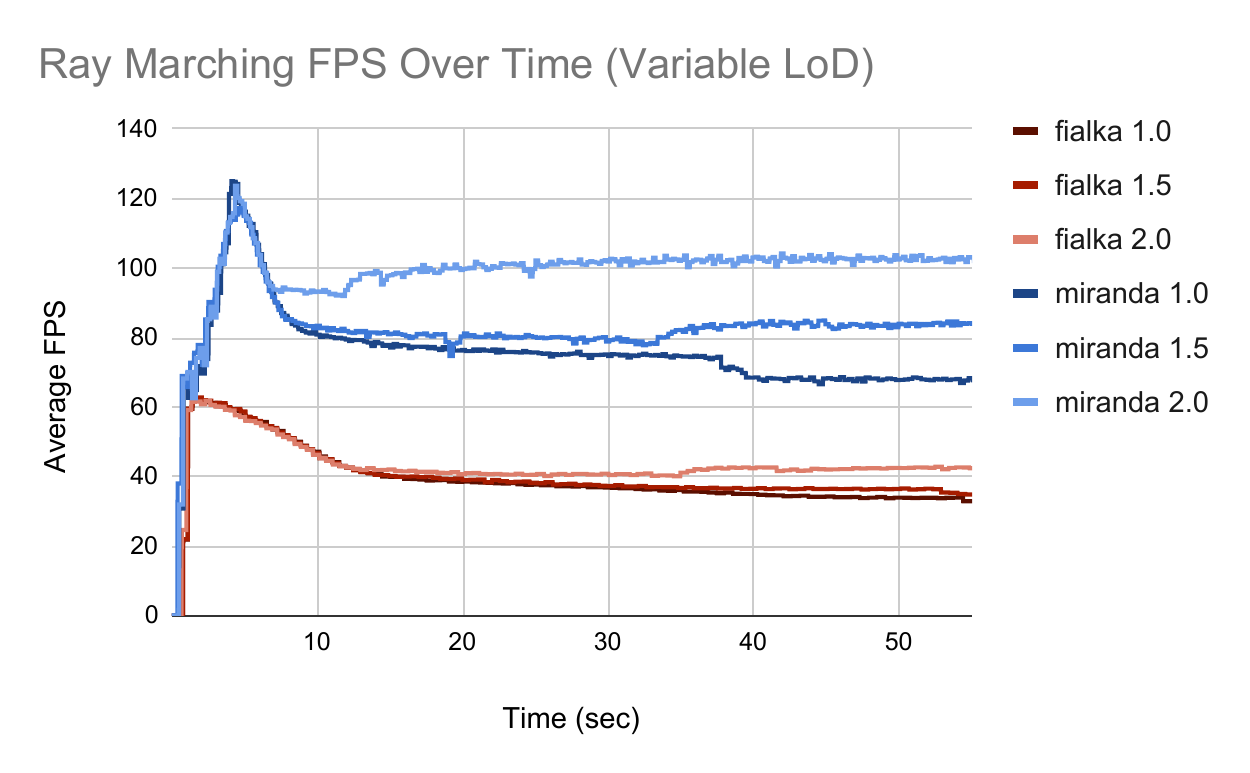}
    \end{subfigure}
    \begin{subfigure}[t]{1.0\linewidth}
        \includegraphics[width=\linewidth]{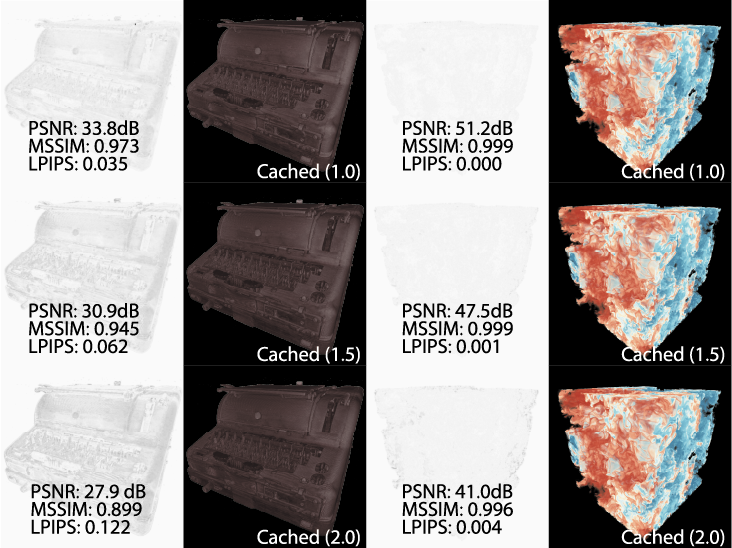}
    \end{subfigure}
    \caption{A performance and quality comparison between a sensitive and resilient dataset to changes in LoD scaling. (Top) A Graph showing the measured FPS for the two datasets at different LoD scales. (Left) Quality measurements of the \textit{Fialka} dataset at increasingly aggressive LoD scaling. (Right) Similar measurements for the \textit{Miranda} dataset which is more resilient to LoD scaling. We see that datasets with more low frequency data are more forgiving at increased LoD and offer higher performance gains at increasing scales.}
    \label{fig:var-lod}
    \vspace{-0.5em}
\end{figure}

\subsection{Rendering Quality}
We calculate all quality metrics in comparison to a pure INR reconstruction of the volume to demonstrate the effects and trade-offs of multi-resolution caching. In general, datasets with primarily low frequency features are the most forgiving at aggressive LoD scales. Looking at \textit{miranda} from our variable LoD demonstration as an example in \cref{fig:var-lod}, even if we double our LoD aggressiveness, pushing interactive rendering performance above 100fps, we still achieve a reconstruction PSNR of 41dB and an MSSIM of 0.996. This along with our results on \textit{heatrelease} point to our pipeline having great effectiveness on visualizing simulated fluid or aerodynamics related data. 

\begin{figure}[h]
    \centering
    \includegraphics[width=\linewidth]{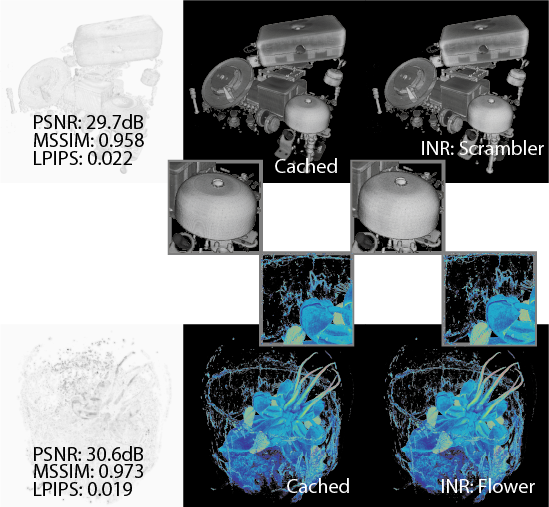}
    \vspace{-1em}
    \caption{The \textit{Scrambler} (Top) and \textit{Flower} (Bottom) datasets show the challenges that flat surfaces (Top) and fine structures (Bottom) pose to reconstruction quality when rendering at higher LoD. Datasets with these features require more conservative LoD scaling to preserve quality and a comparatively larger cache to increase performance.}
    \label{fig:lod-artifacts}
    \vspace{-0.5em}
\end{figure}

On the other hand, we observe two areas where our pipeline struggles; small, high frequency regions and thin, flat surfaces. These regions are the most visibly affected by changes in LoD. Every time the LoD of a brick increases, the distances between each sampled voxel in the brick doubles. This leads to very fine details being entirely skipped at higher LoD. This is most clearly demonstrated in our \textit{flower} dataset where small smudges on the glass dome and details in the roots far away from the camera fade away as shown in \cref{fig:lod-artifacts}. While the affect on surfaces is seen clearly as striations that develop on the typewriter (\textit{fialka}) data in \cref{fig:var-lod}, and on the bell from the scrambler data in \cref{fig:lod-artifacts}.

These issues are most pronounced when the regions are further away from the camera, thus forcing less aggressive LoD settings to be used to preserve overall image fidelity at certain camera angles and distances. We believe that there is potential for future multi-resolution visualization work to identify these kinds of regions and selectively decide LoD. Our current method only takes the sample's distance from the camera into account, which leads to difficulties on some datasets. 

\subsection{Memory Footprint}
\label{sec:memory_footprint}
Since all datasets far exceed the available VRAM, we use the largest, stable GPU cache size possible across all of our up-scaled datasets. We set the grid size to $30\times30\times30$ bricks, with a side length of 40 voxels each. Each brick therefore holds 64,000 voxels for a final cache size of $30^3\times40^3 = 1.728\times10^{9}$ voxels. Our INR always outputs floating point values, so this equates to a GPU cache that is just under 7GB. The remaining roughly 4GB of usage is split between the MRPD, brick loading buffer, macrocell acceleration structure, and various internal buffers used for rendering. The size of the MRPD scales with the data dimensions to accommodate the increased number of bricks. However, as mentioned in \cref{sec:cache-arch}, the MRPD is virtualized for large datasets with layers of cached page tables to limit the structure's memory footprint. The buffer mentioned in \cref{sec:INR-integration} for brick loading adds a static amount of overhead depending on the brick request length. We chose both a brick size of 40 and a request queue length of 40 for this work, resulting in a buffer of size $40^4\times4 = 10.24$MB. The final notable source of memory allocation is with our macro-cell acceleration structure covered in \cref{sec:adaptive-sampling}. The size of each cell is user defined and can be increased on large datasets to lower memory footprint, but even with our default $16\times16\times16$ cell dimensions, a raw dataset the size of our up-scaled ``Miranda'' cube will only produce a structure of $(\frac{4096}{16})^3 = 16.8$ million cells for a final memory footprint of 134MB. Increasing the cell side length on such a volume to 32 or 64 will bring this cost down to just under 17MB and 2MB respectively. Once these parameters are set, our memory usage remains constant throughout rendering. 

\subsection{Hyperparameters}
Our pipeline contains two user-defined parameters that have a notable impact on performance without affecting quality; \textit{brick size} and \textit{maxNumRequests}. We found through testing that a \textit{brick size} of 40, ie. 64,000 voxels gives stable performance on our hardware by limiting the number of cycles the GPU has to wait for each brick to inference. Very large bricks may fill the cache in fewer frames, but the increased computational overhead can lead to worse interactive performance as each brick inference interrupts the GPU for a longer period of time. Conversely, smaller brick sizes will lead to the cache filling more slowly over time, taking longer to reach peak performance at the desired level of quality. This ties into the effects of the second parameter, \textit{maxNumRequests}, which limits the amount of bricks being requested at any given time. The most important detail to keep in mind with this parameter is the behavior once the cache is full. When rendering large data, it is only possible for the cache to represent the entire volume if an aggressive LoD scale is used. In the majority of use cases, cache misses are unavoidable even after rendering has stabilized. This results in a constant stream of brick requests and loading every few frames. Lowering this value will improve peak performance once the cache is full. However, this greatly slows down the rate the cache fills at, limiting interactivity even when the cache is full as it takes longer to adapt to new viewing angles. We chose a request size of 40 to balance interactivity with cache-fill speed in our testing. 

\begin{figure}[h]
    \centering
    \begin{subfigure}[t]{1.0\linewidth}
        \includegraphics[width=\linewidth]{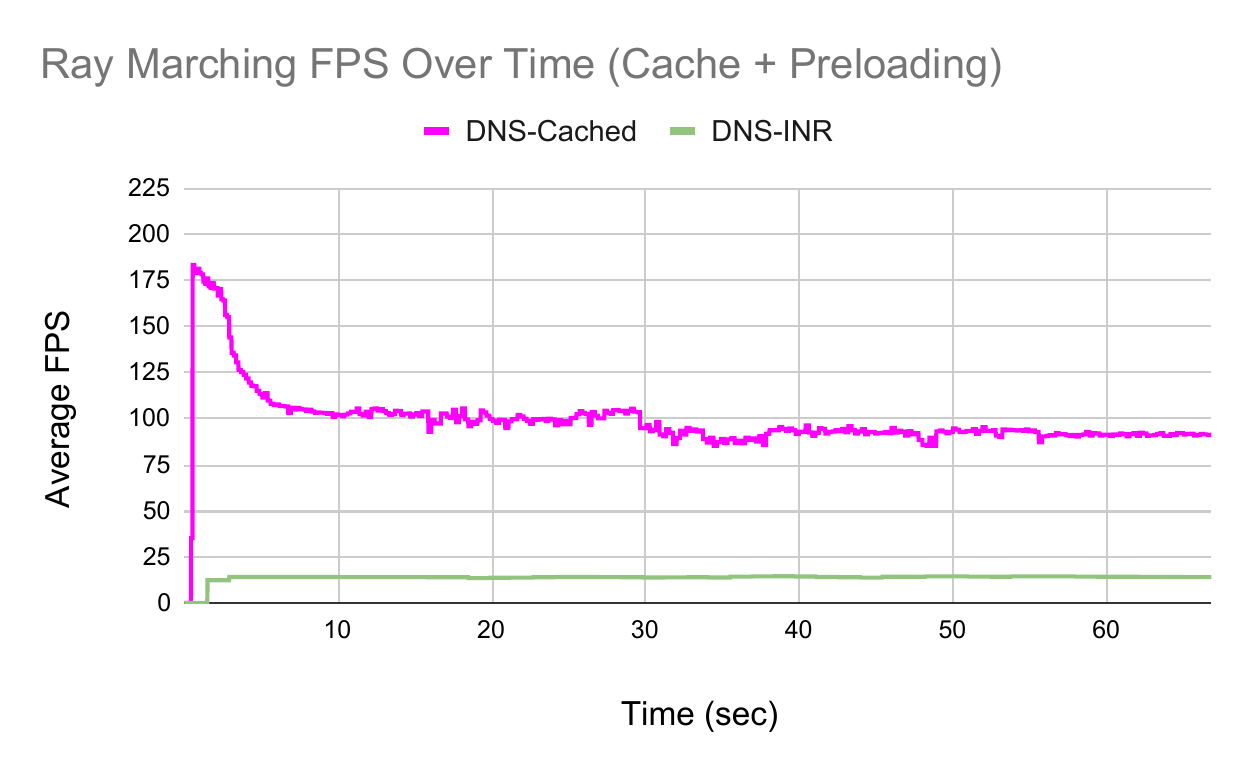}
    \end{subfigure}
    \begin{subfigure}[t]{1.0\linewidth}
        \includegraphics[width=\linewidth]{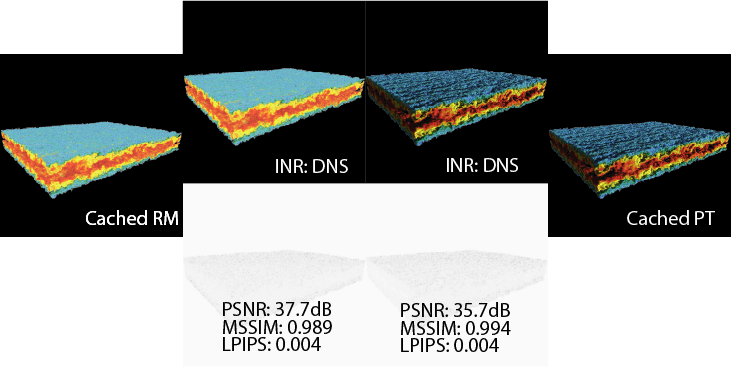}
    \end{subfigure}
    \caption{Both a performance and quality comparison of the rendered DNS data. (Top) plots the rendering FPS of DNS in both network-only and cache-enabled pipelines using our ray marching algorithm. The bottom half of the figure shows our rendered results and quality metrics for both the ray marching (Left) and path tracing (Right) pipelines. These metrics are calculated identically to our previous results.
    }
    \label{fig:dns}
    \vspace{-0.5em}
\end{figure}

\subsection{Case Study}
Our results so far have shown speedups on small and mid-sized datasets which we've up-scaled at render time. We present also a demonstration of our post hoc pipeline on a much larger dataset. This study is conducted on a double-precision, 0.96TB channel flow DNS \cite{dns} dataset. We used an RTX 4090 workstation to compress this data, resulting in a 150MB file with the macro-cell value ranges and model weights, achieving a compression ratio 6444:1 while preserving a PSNR of 35db against the ground truth data. By keeping our model small, we were able to maintain a cache size of $30\times30\times30$, keeping in line with our previous rendering experiments. Following suit, we record a similar performance curve and speedup in \cref{fig:dns} of nearly 6.5$\times$ for the ray marching pipeline, and 2$\times$ for path tracing when compared to our INR-only renderer. Our quality metrics are also in line with previous results despite the larger scale, pointing to the high scalability of our cached pipeline on large data. This, in fact, suggests that our caching method has even greater relative performance at larger scales as the sample count increases and the network overhead becomes more pronounced. 

\section{Discussion and Future Work}
The evaluation results show that our method is effective at reducing the cost of repeated network inference during INR-based rendering. However, our method is not without limitations. We identify future research directions that could enhance our work. 

\noindent \textbf{Memory Footprint.}
We discuss the implications for the memory footprint of our method in \cref{sec:memory_footprint}. In the current version, we rely of full-precision floating point data to maintain our cache which leaves opportunities for further optimization. One possible extension to this could be the use of half-precision floats for retrieval and storage which could cut memory requirements of certain parts of the pipeline in half. Furthermore, there is potential for mixed-precision data encoding based on LoD and user defined error-bounds.

\noindent \textbf{LoD Selection.} Our current approach determines a sample’s Level of Detail solely based on its distance from the camera, which can lead to quality degradation for datasets with fine structures located far from the viewpoint. A more context-aware selection method could better preserve fidelity in these regions while still benefiting from a lower LoD in less critical areas; for instance, leveraging our macro-cell structure---which provides the data value range near each sample---or incorporating a dedicated surface detection technique, could enable dynamic selection of more appropriate LoDs. Additionally, combining multiple heuristics for LoD selection might yield further performance gains by rendering at finer LoDs in sensitive regions and preserving detail at greater distances. 

\noindent \textbf{Cache Representations.} MLP-based implicit neural representations are inherently expensive to infer, with much research focused on hardware acceleration and reducing the size of MLP layers to boost performance. Recent work on 3D Gaussian representations~\cite{kerbl20233dgaussiansplattingrealtime} has delivered incredible performance and memory efficiency gains for NeRF rendering applications, and may have similar potential for INR applications as well.

\section{Conclusion}
We introduce a novel caching technique that significantly enhances the rendering performance of implicit neural representations (INRs) for scientific visualization of large-scale data. Our approach seamlessly integrates state-of-the-art INR compression technologies with multi-resolution GPU caching methods, preserving the memory and performance benefits of both. Agnostic to the underlying INR architecture, our method achieves up to a 5-fold improvement in rendering performance compared to directly rendering INRs, thereby enhancing the usability of interactive scientific visualization systems. 
Our work underscores the potential of integrating INR-based data reduction techniques into extreme-scale, interactive scientific visualization workflows.
We hope our findings inspire further research in high-performance volume rendering and help establish neural methods as a foundational element in scientific visualization.

\section{Acknowledgments}

This research is supported in part by the Intel oneAPI Center of Excellence and National Science Foundation via grant no. IIS-2427770. 
We thank Jonathan Sarton, Welcome Alexandre-Barff, and Laurent Lucas for their valuable insights into their GPU caching methods that helped us improve this work.

\bibliographystyle{eg-alpha-doi}
\bibliography{cINR-egpgv25}

\end{document}